\documentclass[12pt,showpacs,showkeys,preprintnumbers,nofootinbib]{revtex4-2}

\usepackage{amssymb,amsmath,graphics,graphicx}

\usepackage[dvipsnames]{xcolor}

\begin{document}

\newcommand{\pderiv}[2]{\frac{\partial #1}{\partial #2}}
\newcommand{\deriv}[2]{\frac{d #1}{d #2}}

\title{After 2018 Bolsonaro victory, is a 2022 remake feasible?}

\author{Nuno Crokidakis $^{1}$}
\thanks{E-mail: nuno@mail.if.uff.br}

\author{Serge Galam $^{2}$}
\thanks{E-mail: serge.galam@sciencespo.fr}

\affiliation{$^{1}$ Instituto de F\'{\i}sica, Universidade Federal Fluminense, Niter\'oi, Rio de Janeiro, Brazil \\
$^{2}$ CEVIPOF - Centre for Political Research, Sciences Po and CNRS, Paris, France
}

\date{\today}

\begin{abstract}
\noindent
We propose a contagion model to describe the evolution of the political voting trends in Brazil after the dictatorship from 1985 to nowadays. We consider a fully-connected population divided in two voting groups, left and right. Each group includes three kinds of agents, sensitives, inflexibles and radicals. While sensitives may shift their left or right voting, inflexibles and radicals do not. Excluding political interactions with radicals, the model has one interaction parameter, and we found the values which reproduce all the voting outcomes of past presidential elections from 1989 till 2018. As an alternative approach to explain 2018 election, we found that adding an empty third voting group overlapping left and right, the populist voting group, can also yield Bolsonaro 2018 victory. The initial filling of the populist group is triggered by the breaking of the interaction barrier between sensitives and radicals, which had prevailed for decades. Within each voting group, some of the interacting pairs sensitive/radical recast the sensitive into a populist. This status change from sensitive to populist allows the populist to interact with sensitives of both left and right, thus creating an additional source of filling of the populist group. The process is shown to yield the 2018 $55\%$ for Bolsonaro. From the 2018 distribution of votes, we evaluate the parameter changes which recover the recent 2021 poll yielding a victory of Lula against Bolsonaro at a score of 64/36. Thus, we compare the results for the two scenarios, with and without populists, for 2018 election and 2021 poll. Regarding the two possible scenarios for the 2018 Bolsonaro victory, we discuss about a hypothetic Lula/Bolsonaro 2022 second round voting and the requirements to yield either Lula or Bolsonaro victory in 2022. Both scenarios are feasible.

\end{abstract}


\maketitle

\section{Introduction}

The rise of populism has been observed in many democratic countries in the last years \cite{team_populism1,team_populism2,salgado}.  Two most emblematic cases are the elections of Donald Trump in the 2016 US election \cite{winberg} and Jair Bolsonaro in the 2018 Brazilian election \cite{bolsonaro_populist}. In parallel to those successful victories of populist leaders, far-right has grown in a series of Western European countries including France, Germany, Italy and Spain \cite{qz}.

Most of these drastic changes in electoral landscapes were not anticipated by polls and pandits, taken them by surprise. The underlying mechanisms driving the opinion dynamics towards populism and its connection to extreme political views on both the right and the left are still not well understood. In particular the various amplitudes from one country to another are distributed over quite a large spectrum of values.

A recent study investigated the issue in Germany analyzing the changes in political narratives \cite{populism_germany}. Here we tackle the problem from the viewpoint of opinion dynamics studied within the frame of sociophysics. A great deal of works have been published along this approach \cite{sznajd_left_right, rmp}. 

Several studies  related to politics and elections were published in the last years \cite{Bottcher_plos,Gracia,Araripe,galam_trump}. They use discrete variables or continuous ones. Among them, simple contagion models were considered to study the evolution of distinct social processes, like tax evasion \cite{ijmpc_rafael}, cooperation \cite{lima2014}, rumor spreading \cite{moreno}, epidemic spreading \cite{satorras_rmp}, radicalization phenomena \cite{galam_javarone}, ideological conflicts \cite{marvel}, corruption \cite{jorge}, juvenile crimes \cite{lee}, obesity \cite{ejima}. 
 
In this work we consider a fully-connected population composed of $N$ individuals. Agents are labelled in terms of political affiliations, either left or right. Each group includes three kinds of agents, sensitives, inflexibles and radicals adding to a total of six subgroups. These subgroups are denoted sensitive, inflexible, radical. Their respective proportions are $l$, $ll$, $L$ for the left, and $r$, $rr$, $R$ for the right. 

While inflexibles and radicals do not shift their political affiliations, sensitives are liable to shift to the opposite side via some specific pair interactions. Accordingly, while proportions $ll$, $L$, $rr$, $R$ are given and constants, proportions $l$ and $r$ are functions of time. What makes an individual to become radical or inflexible is out the scope of the present work. We thus have the conservation condition,
\begin{equation}\label{eq1}
l(t)+r(t)+L+ll+R+rr=1 .
\end{equation}

We first apply the model to explain the voting outcome for 1989 to 2014 Brazilian elections. After that, we introduce the populists in the model, in order to give an explanation for the Bolsonaro victory in 2018 election. We also consider the recent 2021 poll regarding a possible Lula vs Bolsonaro in a second round in the next year presidential election. We also present an alternative explanation for the Bolsonaro victory in 2018 election, based on a single varying parameter without populists. This alternative scenario can reproduce all election outcomes (1989 - 2018) and the 2021 poll with a single parameter. For 2018 election and 2021 poll, we compare the results of the two scenarios, with and without populists. Considering 2018 election, a recent work by social scientists that discusses the actual impact of populist attitudes in the election of Bolsonaro is found in agreement with the finding of our mathematical model \cite{castanho2022}. Finally, regarding the two possible scenarios for the 2018 Bolsonaro victory, we discuss about a hypothetic Lula/Bolsonaro 2022 second round voting and the requirements to yield either Lula or Bolsonaro victory in 2022.


\section{Competition between left and right over sensitives}

During last decades since last World War II, with few exceptions radicals have been excluded from social discussions about voting choices among people from left and right. On both sides, left and right, non radicals refused to discuss with their side-extremists often labelled  ``fascists" for the right and ``communists" for the left. Discussions were restricted between inflexibles and sensitives. During these exchanges, sensitives could eventually shift side. The various interactive configurations with their respective output are,
\begin{eqnarray} \label{eq2}
l + r& \stackrel{k}{\rightarrow} & l + l , \\ \label{eq3}
l + r & \stackrel{1-k}{\rightarrow} & r + r , \\ \label{eq4}
ll + r& \stackrel{k}{\rightarrow} & ll + l  , \\ \label{eq5}
ll + r& \stackrel{1-k}{\rightarrow} & ll + r  , \\ \label{eq6}
l + rr & \stackrel{1-k}{\rightarrow} & r + rr  , \\ \label{eq7}
l + rr & \stackrel{k}{\rightarrow} & l + rr  
\end{eqnarray}
where $ 0 \leq k \leq 1$. For simplicity, we have assumed equal shift probabilities. For each case, sensitives either shift side or keep on their current stance.  The rate $k$ measures the propensity:
\begin{itemize}
\item for a left sensitive to convince a right sensitive and $(1-k)$ the reverse.
\item  for a left inflexible to convince a right sensitive while $(1-k)$ is the capacity of a sensitive right to resist this outcome. 
\item for a left sensitive to resist a right inflexible and $(1-k)$ the reverse.  
\end{itemize}
We  also consider equal fixed proportions $f=L=ll=R=rr$ for left and right inflexibles and radicals. At this stage it is worth  stressing that our goal focuses on getting a minimal coherent frame that yields all the Brazilian elections' results to address the question of the impact of populism if any in Bolsonaro 2018 victory.

The corresponding time evolution of $l(t)$ and $r(t)$ are given by,
\begin{eqnarray} \label{eq8}
\frac{dl(t)}{dt} & = & -(1-2k)\,l(t)\,r(t) + k\,f\,r(t) - (1-k)\,f\,l(t) ,\\ \label{eq9}
\frac{dr(t)}{dt} & = & (1-2k)\,l(t)\,r(t) - k\,f\,r(t) + (1-k)\,f\,l(t) ,
\end{eqnarray}
where $\frac{dl(t)}{dt} =-\frac{dr(t)}{dt}$ as expected from the fact that the dynamics of affiliation occurs only within the group of sensitives with $l(t) + r(t) = 1-ll-L-rr-R= 1 - 4f= constant$.

In such a case, solving the equation $\frac{dl}{dt} = 0$ yields the steady state,
\begin{equation} \label{eq10}
l = \frac{k\,f\,r}{(1-k)\,f - (2k-1)\,r}    
\end{equation}
\noindent
where $l=l(t\to\infty)$ and $r=r(t\to\infty)$ are the stationary values of $l(t)$ and $r(t)$, respectively. From Eq. \eqref{eq1}, substituting $r=1-l-4f$ into Eq. \eqref{eq10} yields a second order polynomial for $l$, 
\begin{equation} \label{eq2l}
Al^{2} + Bl + C = 0 ,
\end{equation}
where
\begin{eqnarray} \label{eq11}
A &  \equiv  & -(1-2k) ,\\ \label{eq12}
B &  \equiv  & (1-k)\,f + (1-2k)\,(1-4f) + k\,f ,\\ \label{eq13}
C &  \equiv  & -k\,(1-4f)\,f  ,
\end{eqnarray}
From Eq. \eqref{eq2l}  we obtain 2 solutions,
\begin{equation}\label{eq15}
l = \frac{-[(1-k)\,f + (1-2k)\,4f+k\,f] \pm \sqrt{\Delta}}{2\,(2k-1)} .
\end{equation}
\noindent
where $\Delta=B^{2} - 4AC$, with $A$, $B$ and $C$ given by Eqs. \eqref{eq11} - \eqref{eq13}. Numerically, we found that the solution with the \textit{plus} signal is the relevant one giving $l>0$. The stationary value for the sensitive right fraction $r$ can be obtained from the normalization condition (Eq. \eqref{eq1}), i.e., $r=1-l-4f$.

In particular, the case $k=1/2$ of symmetry balance between shift from sensitive interactions yields,
\begin{eqnarray} \label{eq18}
l & = & \frac{(1-4f)}{2} \\ \label{eq19} 
r & = & \frac{(1-4f)}{2} .
\end{eqnarray}


\section{Applying the model to Brazilian elections}

After a long dictatorship (1964 - 1985) a democratic system has been installed based on a two rounds' presidential elections. Voting is mandatory although penalty fee for not voting is moderate \footnote{The value vary between $R\$ 1.05$ and $R\$ 3.51$ in Brazilian currency (Reals, where $R\$ 1$ corresponds to $0.18$ US$\$$).}. In 1989, the first president to be elected has been the center-right candidate Fernando Collor de Mello. Following elections in 1994 and 1998 were won by the right-wing candidate Fernando Henrique Cardoso at first rounds. The left-wing candidate Luis In\'acio Lula da Silva was elected twice in a row at second rounds in 2002 and 2006. His successor, Dilma Roussef, was also elected twice in a row at second rounds in 2010 and 2014. Finally, the far-right candidate Jair Bolsonaro was elected in 2018 at second round \cite{election_results}.

For the above-mentioned elections, six had a second round (1989, 2002, 2006, 2010, 2014, 2018) and two (1994, 1998) had the president elected at the first round \cite{election_results}. Respective voting outcomes are:

\begin{itemize}

\item $1989: 0.53$ (right-wing candidate) and $0.47$ (left-wing candidate)

\item $1994: 0.61$ (right-wing candidate) and $0.39$ (left-wing candidate)
  
\item $1998: 0.57$ (right-wing candidate) and $0.43$ (left-wing candidate)

\item $2002: 0.39$ (right-wing candidate) and $0.61$ (left-wing candidate)

\item $2006: 0.39$ (right-wing candidate) and $0.61$ (left-wing candidate)

\item $2010: 0.44$ (right-wing candidate) and $0.56$ (left-wing candidate)

\item $2014: 0.48$ (right-wing candidate) and $0.52$ (left-wing candidate)

\item $2018: 0.55$ (right-wing candidate) and $0.45$ (left-wing candidate)

\end{itemize}

\begin{figure}[t]
\begin{center}
\vspace{6mm}
\includegraphics[width=0.45\textwidth,angle=0]{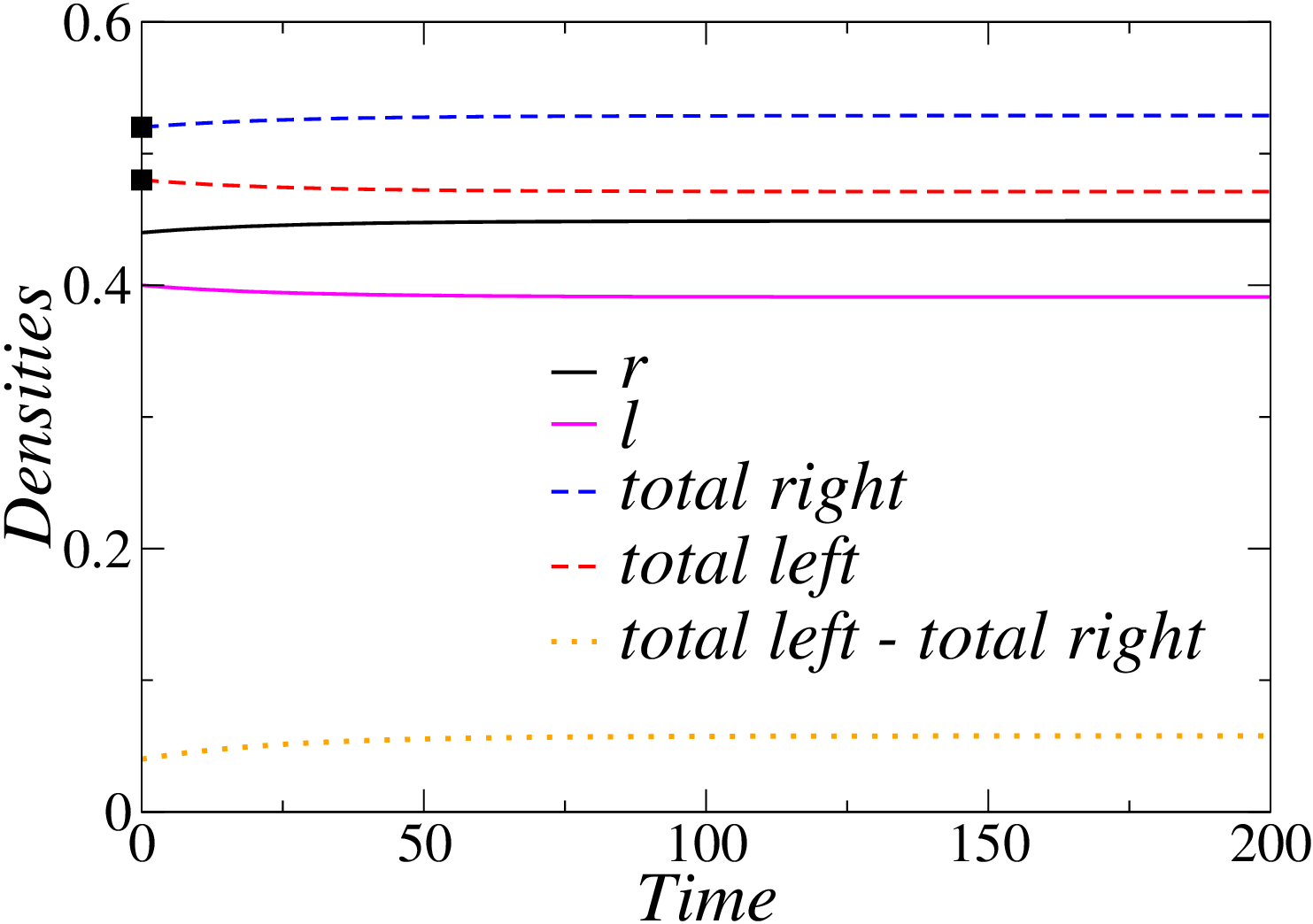}
\includegraphics[width=0.45\textwidth,angle=0]{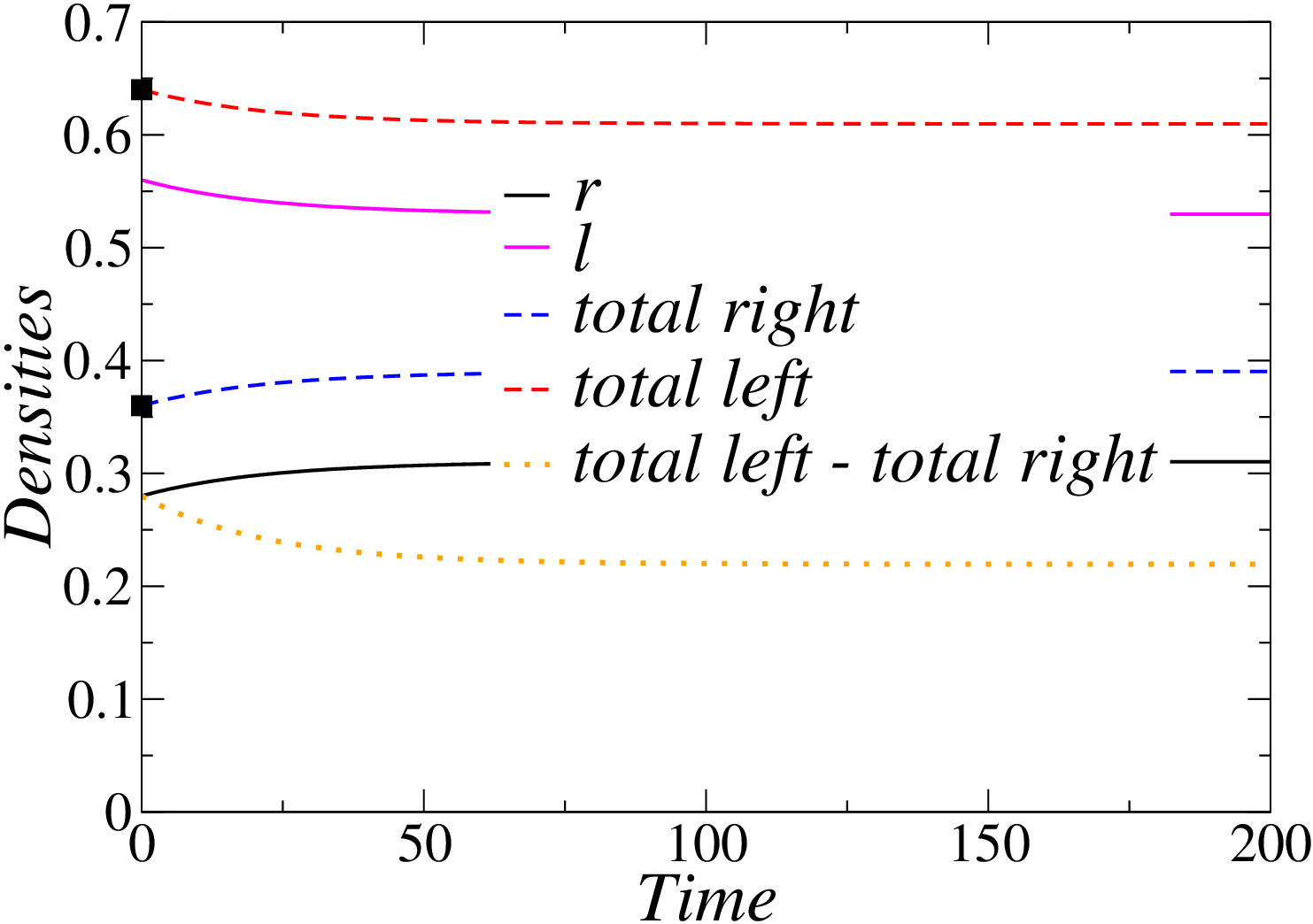}
\\
\vspace{0.5cm}
\includegraphics[width=0.45\textwidth,angle=0]{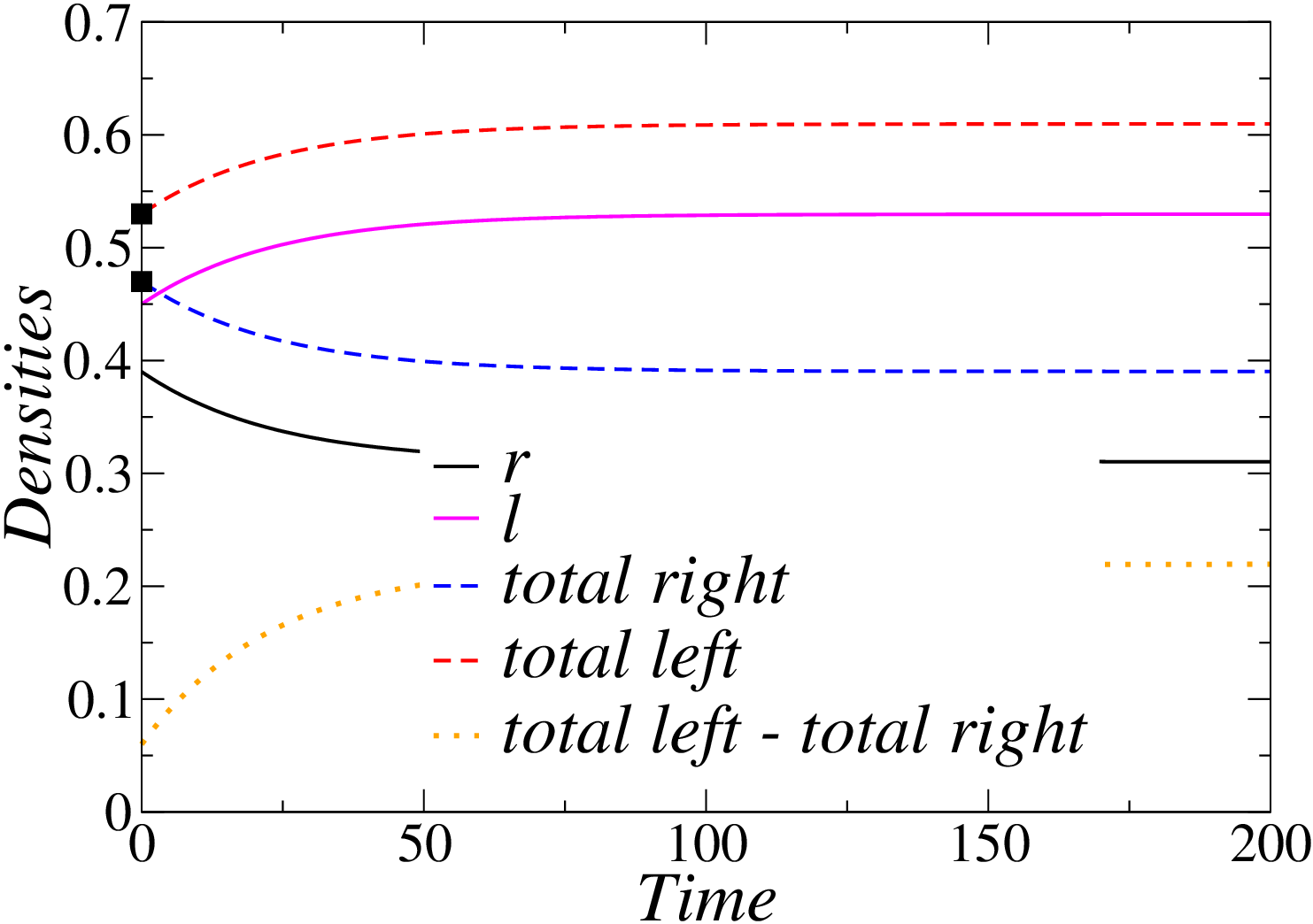}
\end{center}
\caption{(Color online) Time evolution of the densities of sensitives ($l$ and $r$), total right ($r+rr+R$), total left ($l+ll+L$) and the difference $total\,\, left - total\,\, right$. (a: upper left) Simulation of 1989 election. (b: upper right) Simulation of 2002 election. (c: lower) Simulation of 2006 election. The black squares indicate the initial conditions, defined by polls \cite{polls}: (a) $l(0)=0.40, r(0)=0.44$; (b) $l(0)=0.56, r(0)=0.28$; (c) $l(0)=0.45, r(0)=0.39$. We took $f=0.04$ with the value of $k$ for each election given in Table \ref{Tab1}.} 
\label{fig1}
\end{figure}

\begin{figure}[t]
\begin{center}
\vspace{6mm}
\includegraphics[width=0.45\textwidth,angle=0]{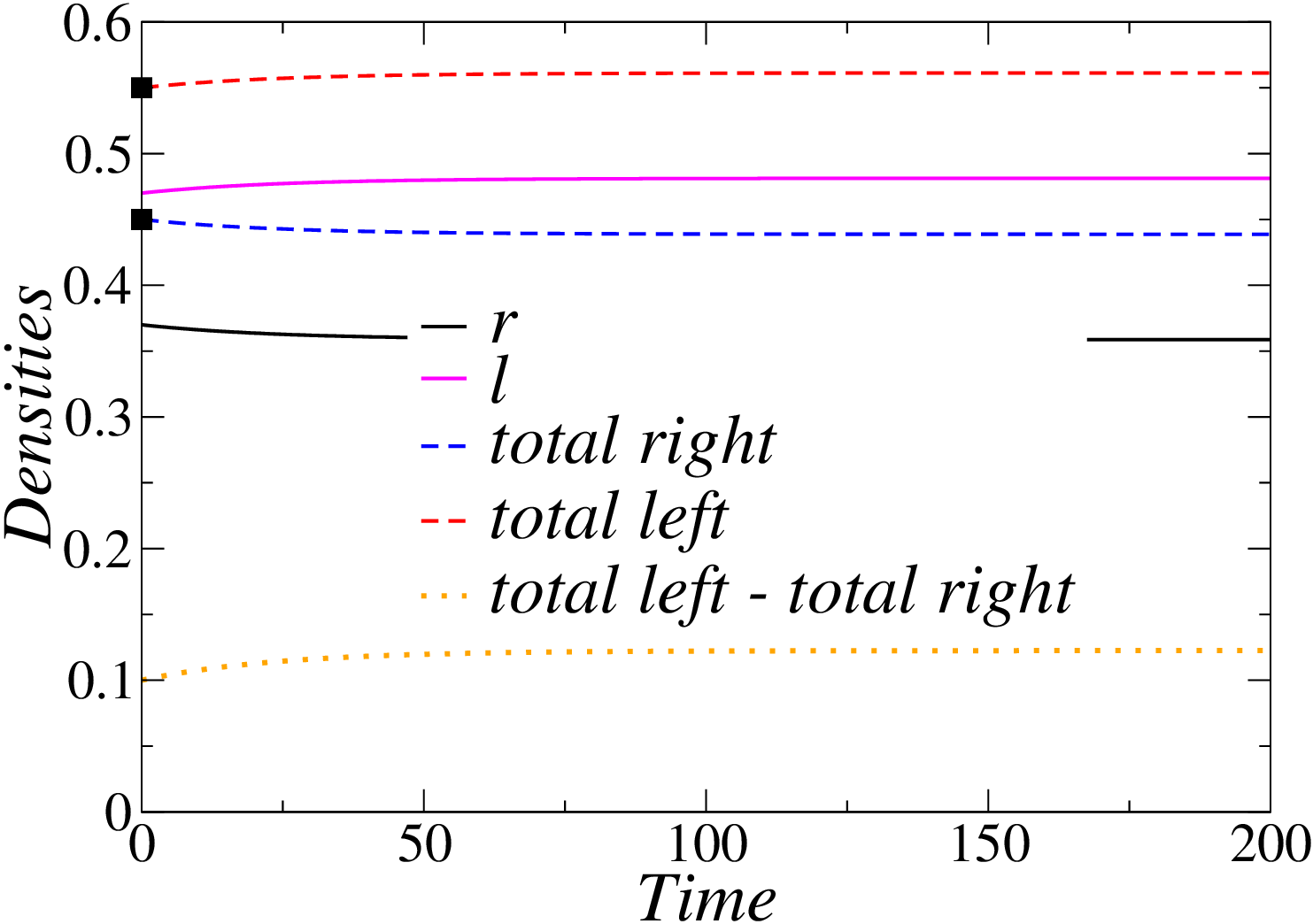}
\includegraphics[width=0.45\textwidth,angle=0]{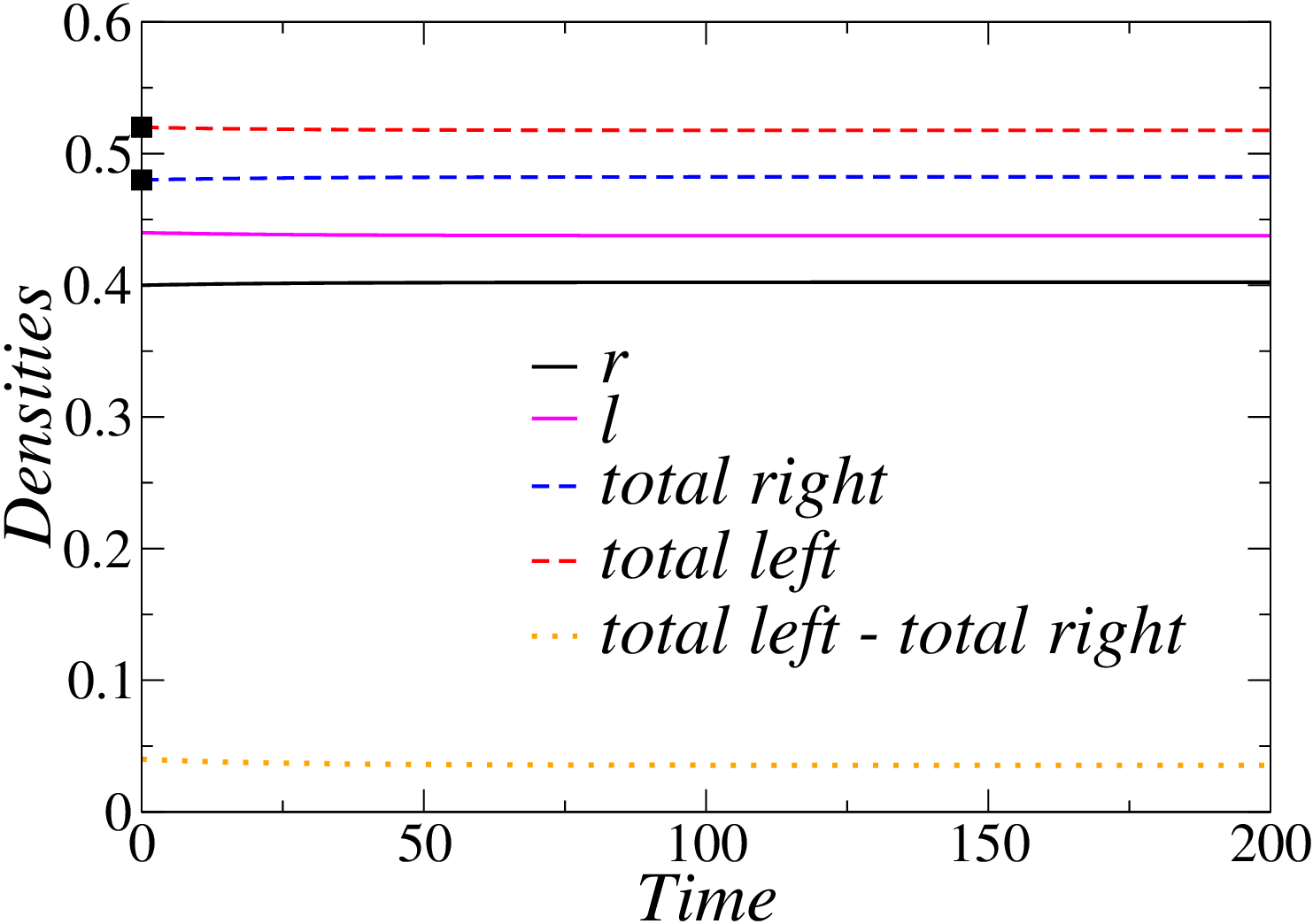}
\\
\vspace{0.5cm}
\includegraphics[width=0.45\textwidth,angle=0]{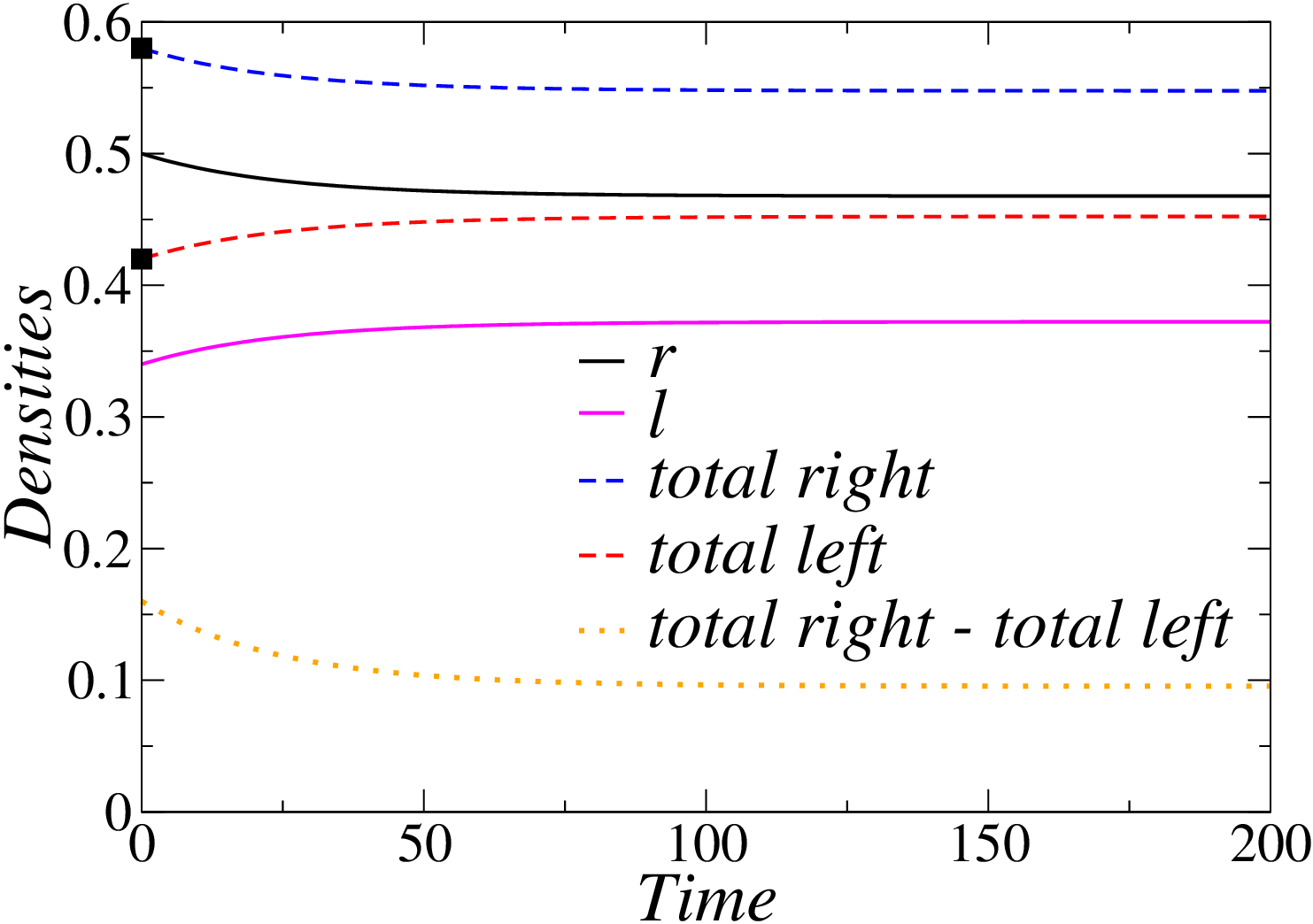}
\end{center}
\caption{(Color online) Time evolution of the densities of sensitives ($l$ and $r$), total right ($r+rr+R$), total left ($l+ll+L$) and the difference $total\,\, left - total\,\, right$. (a: upper left) Simulation of 2010 election. (b: upper right) Simulation of 2014 election. (c: lower) Simulation of 2018 election.  The black squares indicate the initial conditions, defined by polls \cite{polls}: $l(0)=0.47, r(0)=0.37$; (b) $l(0)=0.44, r(0)=0.40$ and (c) $l(0)=0.34, r(0)=0.50$. We took $f=0.04$ with the value of $k$ for each election given in Table \ref{Tab1}.}
\label{fig1_2}
\end{figure}

Above results consider only the valid votes, i.e., they do not take into account null votes or abstentions. Moreover, for the first round 1994 election, the value $0.61$ for the right-wing candidate is obtained summing outcomes of the actual right-wing winner ($\approx 0.54$) and the third candidate also right-wing ($\approx 0.07$). Similarly for the first round 1998 election, we added the outcomes of second ($\approx 0.32$ ) and third ($\approx 0.11$) candidates both being left-wing to yield $0.43$ for the left-wing candidate.

For all those eight elections, fixing $f$ and using  Eqs. \eqref{eq15} and \eqref{eq1} we are able to find appropriate values of $k$ producing steady-state values for total left$=l+ll+L=l+2f$ and total right$=r+rr+R=r+2f$, which are identical to the actual voting outcomes. 

For the next results, we chose $f=0.04$. For such value, we can recover all the voting outcomes as a function of the single parameter $k$. The respective values of $k$ compatible with available data \cite{election_results} are given in Table \ref{Tab1}. We chose $f=0.04$ since it yields a reasonable value for the total density of inflexibles and radicals, namely a bit less than $20\% \,(4f=4\times 0.04 = 0.16)$. Moreover, we found that increasing the total amount to about a quarter has very little effect on  the values of $k$.

\begin{table*}[tbp]
\begin{center}
\vspace{0.1cm}
\renewcommand\arraystretch{1.3} 
\begin{tabular}{|c|c|}
\hline
$Election$ & $k$     \\ \hline
1989 & 0.497  \\
1994 & 0.488  \\
1998 & 0.493  \\
2002 & 0.513  \\ 
2006 & 0.513  \\ 
2010 & 0.507  \\ 
2014 & 0.502   \\
2018 & 0.495  \\
\hline
\end{tabular}%
\end{center}
\caption{Values of $k$ found to match the elections' results from 1989 to 2018 considering the model without populists, for fixed $f=L=ll+R=rr=0.04$.}
\label{Tab1}
\end{table*}

Figures \ref{fig1} and \ref{fig1_2} exhibit the various dynamics obtained from the numerical integration of Eqs. \eqref{eq8} and \eqref{eq9} using the sets of parameters given in Table \ref{Tab1}. In each figure we plot the densities of sensitives $l$ and $r$, the total densities of left ($l+ll+L$) and right ($r+rr+R$) supporters, and the difference $total\,\, left - total\,\, right$. The initial conditions are represented by the black squares, and they were obtained by polls \cite{polls}. The corresponding values $l(0)$ and $r(0)$ for the initial conditions are given in the captions of Figures \ref{fig1} and \ref{fig1_2}.

The model reproduces both victories of ring-wing and left-wing, depending on the value of $k$. From the values of $k$, even taking into account that the radicals do not interact with the other two classes (sensitives and inflexibles), the presence of such radicals is important.


\section{The emergence of populism: An alternative explanation for 2018 Bolsonaro victory}

\subsection{Bolsonaro and Populism}

In 2018, the populist Jair Bolsonaro was elected at second round in the Brazilian presidential election. But why is Bolsonaro considered a populist? We can adopt the criteria of the Team Populism \cite{team_pop_page}. The rubric produced by Team Populism grades the politicians speeches on a scale that goes from $0$ to $2$: $0$ is classified as ``not populist'', $0.5$ as ``somewhat populist'', $1.0$ as ``populist'', $1.5$ as ``very populist'' and leaves $2.0$ open for what they call ``perfect populist'' \cite{bolsonaro_populist}. As stated in \cite{bolsonaro_populist}, unlike past presidents Dilma Rousseff, Lula and Fernando Henrique Cardoso, Bolsonaro articulates stronger populist elements and uses them more often. His predecessors' average scores vary from $0$ to $0.3$, not enough to be considered populist. Based on Bolsonaro's speechs, the Team Populism obtained a grade $0.88$ \cite{bolsonaro_populist}.

Thus, we can make a basic and simple hypothesis to include populists among agents. We assume that populists are former sensitive agents from both left and right, who had interacted with their respective radicals. Populists form a seventh subgroup of agents denoted $p(t)$. In addition, once they are created, populists do interact with sensitives from both left and right. The proportion of populists is thus a function of time and sensitives. The normalization condition given by Eq. \eqref{eq1} becomes,
\begin{equation}\label{eq20}
l(t)+r(t)+p(t)+4f=1 .
\end{equation}
In addition to the social interactions given by Eqs. \eqref{eq2}-\eqref{eq7}, we have,
\begin{eqnarray} \label{eq21}
L + l& \stackrel{\alpha}{\rightarrow} & L + p  ,\\ \label{eq22} 
L + l& \stackrel{1-\alpha}{\rightarrow} & L + l  ,\\ \label{eq23} 
R + r& \stackrel{\alpha}{\rightarrow} & R + p  ,\\ \label{eq24} 
R + r& \stackrel{1-\alpha}{\rightarrow} & R + r  ,\\ \label{eq25} 
l + p& \stackrel{\alpha}{\rightarrow} & p + p  ,\\ \label{eq26} 
l + p& \stackrel{1-\alpha}{\rightarrow} & l + l  ,\\ \label{eq27} 
r + p& \stackrel{\alpha}{\rightarrow} & p + p  ,\\ \label{eq28} 
r + p& \stackrel{1-\alpha}{\rightarrow} & r + r ,\end{eqnarray}

Keeping on with our simplified approach we chose the same parameter $\alpha$ for the various probabilistic shifts. We now have two varying parameters, namely $k$ and $\alpha$. Above interactions rules lead to the evolution equations,

\begin{eqnarray} \nonumber
\frac{dl(t)}{dt} & = & -(1-2k)\,l(t)\,r(t) + k\,f\,r(t) - (1-k)\,f\,l(t) -\alpha\,f\,l(t) \\  \label{eq29} 
&& + (1-2\alpha)\,l(t)\,p(t) ,\\ \nonumber
\frac{dr(t)}{dt} & = & (1-2k)\,l(t)\,r(t) - k\,f\,r(t) + (1-k)\,f\,l(t) - \alpha\,f\,r(t) \\ \label{eq30} 
&& + (1-2\alpha)\,r(t)\,p(t) ,\\ \label{eq31}
\frac{dp(t)}{dt} & = & \alpha\,f\,l(t) + \alpha\,f\,r(t) - (1-2\alpha)\,l(t)\,p(t) - (1-2\alpha)\,r(t)\,p(t) .
\end{eqnarray}

\begin{figure}[t]
\begin{center}
\vspace{6mm}
\includegraphics[width=0.48\textwidth,angle=0]{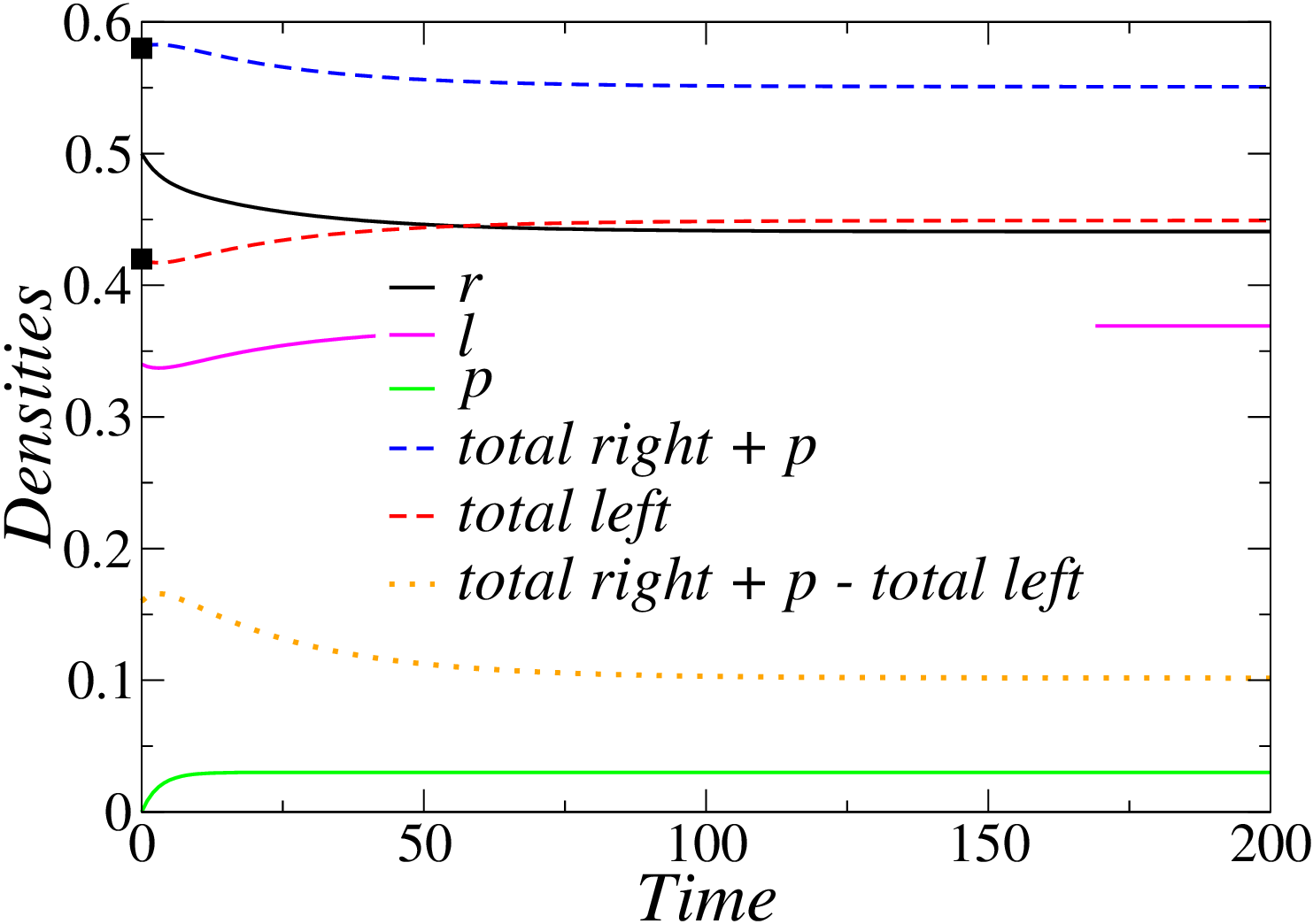}
\includegraphics[width=0.48\textwidth,angle=0]{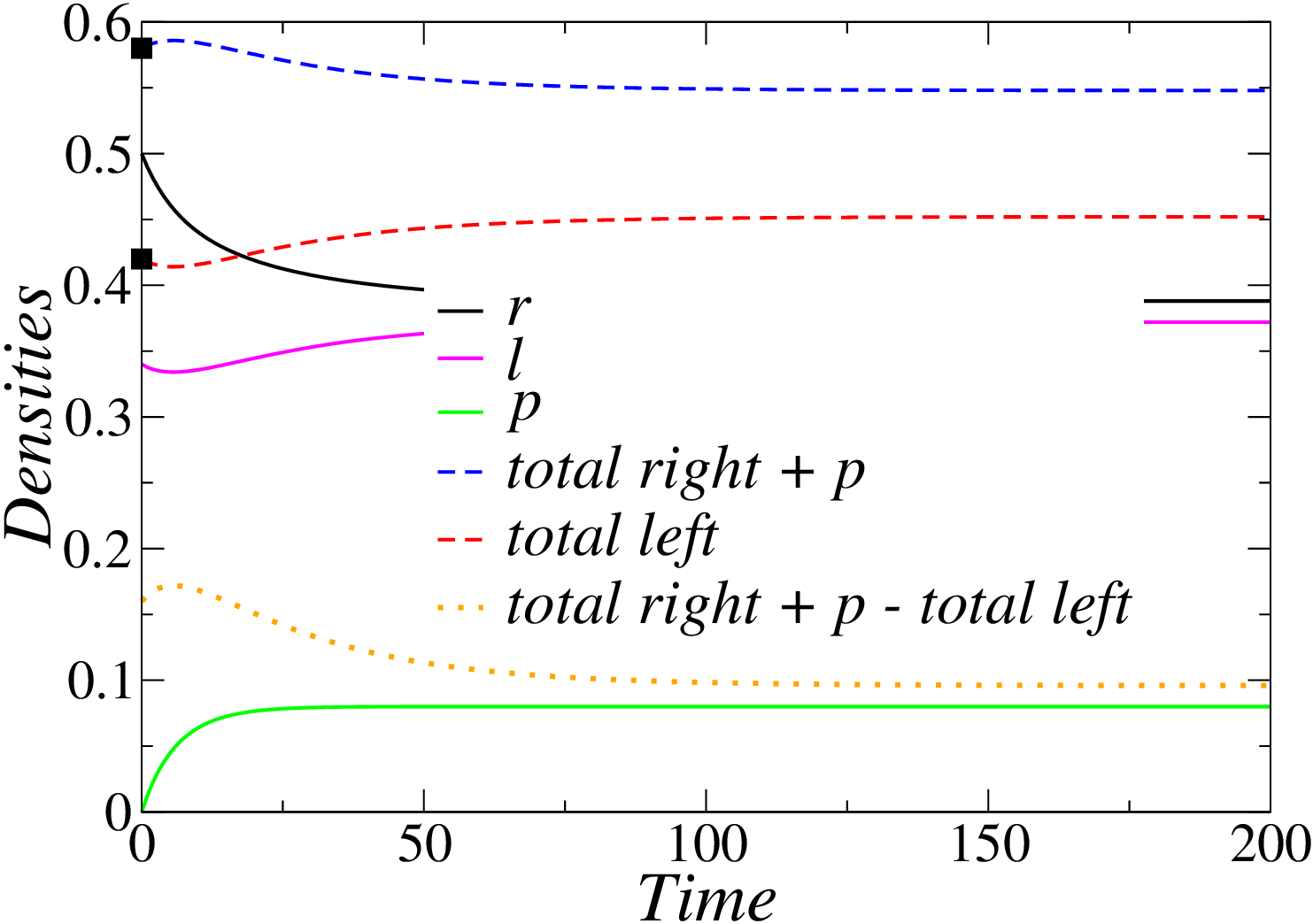}
\end{center}
\caption{(Color online) Time evolution of the densities. These results simulate the evolution of votes in the 2018 election considering the populism coupling. We observe the Bolsonaro victory considering the total right ($r+rr+R$) plus the populists $p$, i.e., for $r+rr+R+p$. (a, left panel:) results $\alpha=0.30$. (b, right panel:) results for $\alpha=0.40$. The black squares in both panels denote the initial conditions, obtained by polls \cite{polls}, i.e., the same values considering in Figure 2 (c), $l(0)=0.34$ and $r(0)=0.50$. We have $f=0.04$ with and the value of $k$ for each election given in Table \ref{Tab2}.}
\label{fig2}
\end{figure}

We performed a numerical integration of the evolution equations \eqref{eq29} - \eqref{eq31} in order to obtain the 2018 election results, namely Bolsonaro received about $55\%$ of the votes and the left-wing candidate Haddad received about $45\%$ of the votes \cite{election_results}. As in the previous section $f=0.04$. In Figure \ref{fig2} we exhibit two graphics showing the time evolution of the densities of interest, for the choices $\alpha=0.30$ (left panel) and $\alpha=0.40$ (right panel). For both graphics, we consider the same initial condition given by polls \cite{polls}, namely $l(0)=0.34$ and $r(0)=0.50$ (the same considered in section III). In addition, we also consider $p(0)=0$. To match the 2018 election's results, we found $k=0.496$ for $\alpha=0.30$ and $k=0.499$ for $\alpha=0.40$. In such a case, based on the above discussion about the Bolsonaro's populism, we considered the votes from the populist compartiment to Bolsonaro, i.e., the total fraction of votes for the right candidate is given by $r+rr+R+p$.

Some comments about the results:

\begin{itemize}
\item
Comparing with the 2014 election, the left-wing candidate support decreased from $0.52$ in 2014 to $0.45$ in 2018. In addition, the right-wing candidate support increased from $0.48$ in 2014 to $0.55$ in 2018.

\item
Based on the model, considering as a case study the value $\alpha=0.40$, we obtained the equilibrium values total left $=0.45$, total right $=0.47$ and $p=0.08$, leading to total right plus populists $= 0.55$. In other words, the fraction of right individuals decreased from $0.48$ in 2014 to $0.47$ in 2018.
  
\item
However, the fraction of populists increased from zero in 2014 to $0.08$ in 2018. Those choices are based on the assumption that right sensitives $r$ turned in populists $p$ from interacting with few right radicals $R$, and once a seed of populists $p$ is formed it sucks directly a part of right sensitives $r$ and also a small part of left sensitives $l$. 
\end{itemize}

\subsection{Comment regarding the 2021 poll about a possible Bolsonaro vs Lula in 2022}

A recent poll was conducted in Brazil in September 2021 about a possible 2022 second round Bolsonaro / Lula (right / left). The results yielded a substantial advantage for Lula with $0.64$ against $0.36$ for Bolsonaro \cite{poll_2021}, a large shift of $0.19$ with respect to the 2018 results.

Within our model, such a shift implies (1) A decrease of the populism support, that was $p=0.08$ for 2018 (for $\alpha=0.40$); (2) A large number of right sensitives $r$ turning into left sensitives $l$. Indeed, an increase in $k$ with a simultaneous decrease of $\alpha$ can produce the poll observed shift of $0.19$ from right to left.


\section{Comparing the two explanations for 2018 Bolsonaro victory}

To explain 2018 Bolsonaro victory, we have first considered the usual competition between left and right for sensitives, without interactions with radicals (model of section II). After that, in section IV.A we introduced the mechanism of interaction between sensitives and radicals to create populists, who then can spread over interacting with sensitives from both left and right. Then, to recover Bolsonaro outcome of $0.55$ we found the equilibrium values $p=0.03$ (for $\alpha=0.30$) and $p=0.08$ (for $\alpha=0.40$) for populists, which, contrary to what was expected for a populist leader, is not that high.

\begin{table*}[tbp]
\begin{center}
\vspace{0.1cm}
\renewcommand\arraystretch{1.3} 
\begin{tabular}{|c|c|}
\hline
$Election$ & $k$     \\ \hline
2018 & 0.495  \\
2018 & 0.496$^{1}$ \\
2018 & 0.499$^{2}$ \\
2021 & 0.516 \\
2021 & 0.520$^{1}$ \\
2021 & 0.528$^{2}$ \\
\hline
\end{tabular}%
\end{center}
\caption{Values of $k$ found to reproduce the elections' results considering the model without populists (model of Section II) and the model with populists (model of Section IV.A) for $\alpha=0.30$ (values with $^{1}$) and $\alpha=0.40$ (values with $^{2}$).}
\label{Tab2}
\end{table*}

\begin{figure}[t]
\begin{center}
\vspace{6mm}
\includegraphics[width=0.48\textwidth,angle=0]{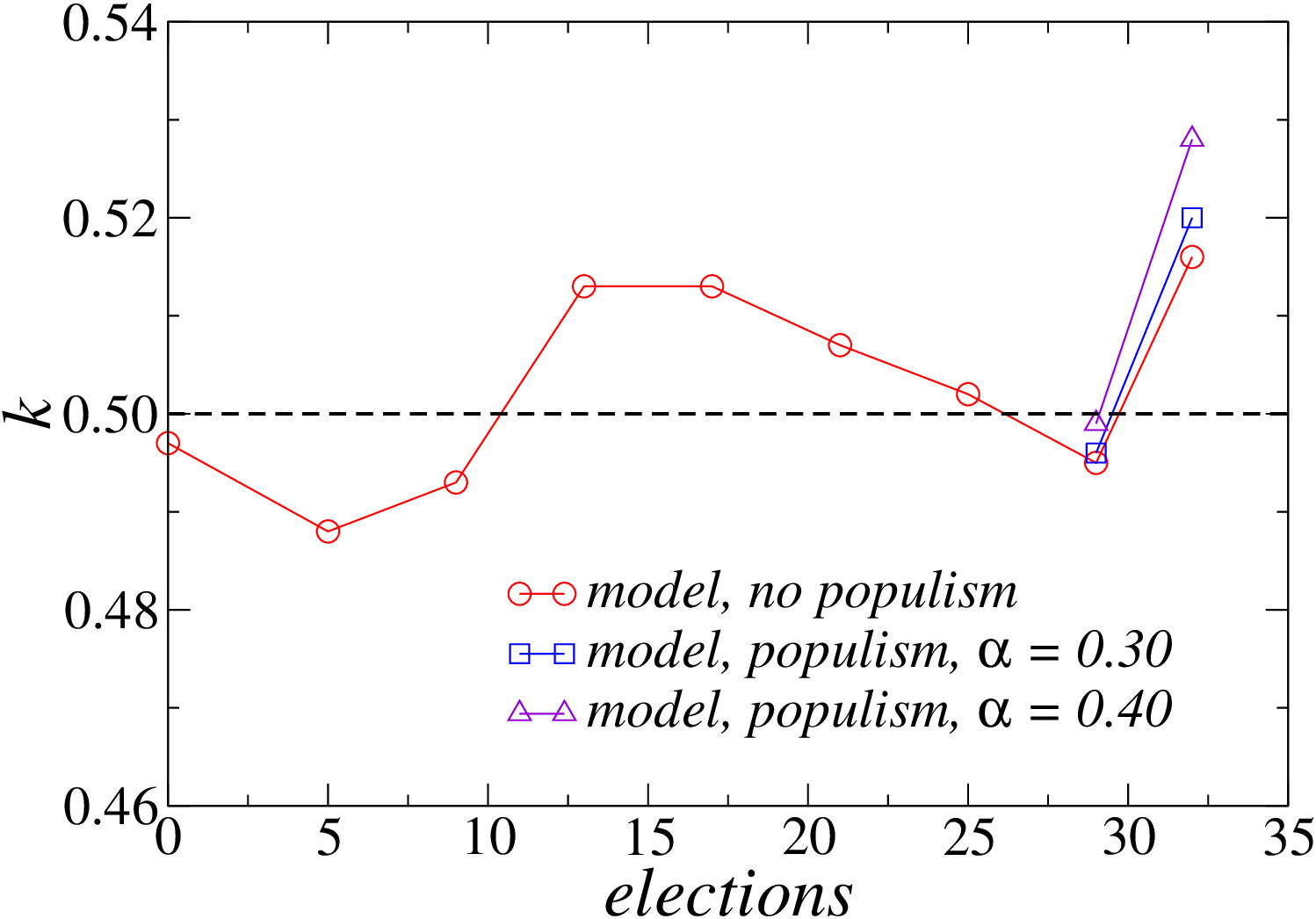}
\includegraphics[width=0.48\textwidth,angle=0]{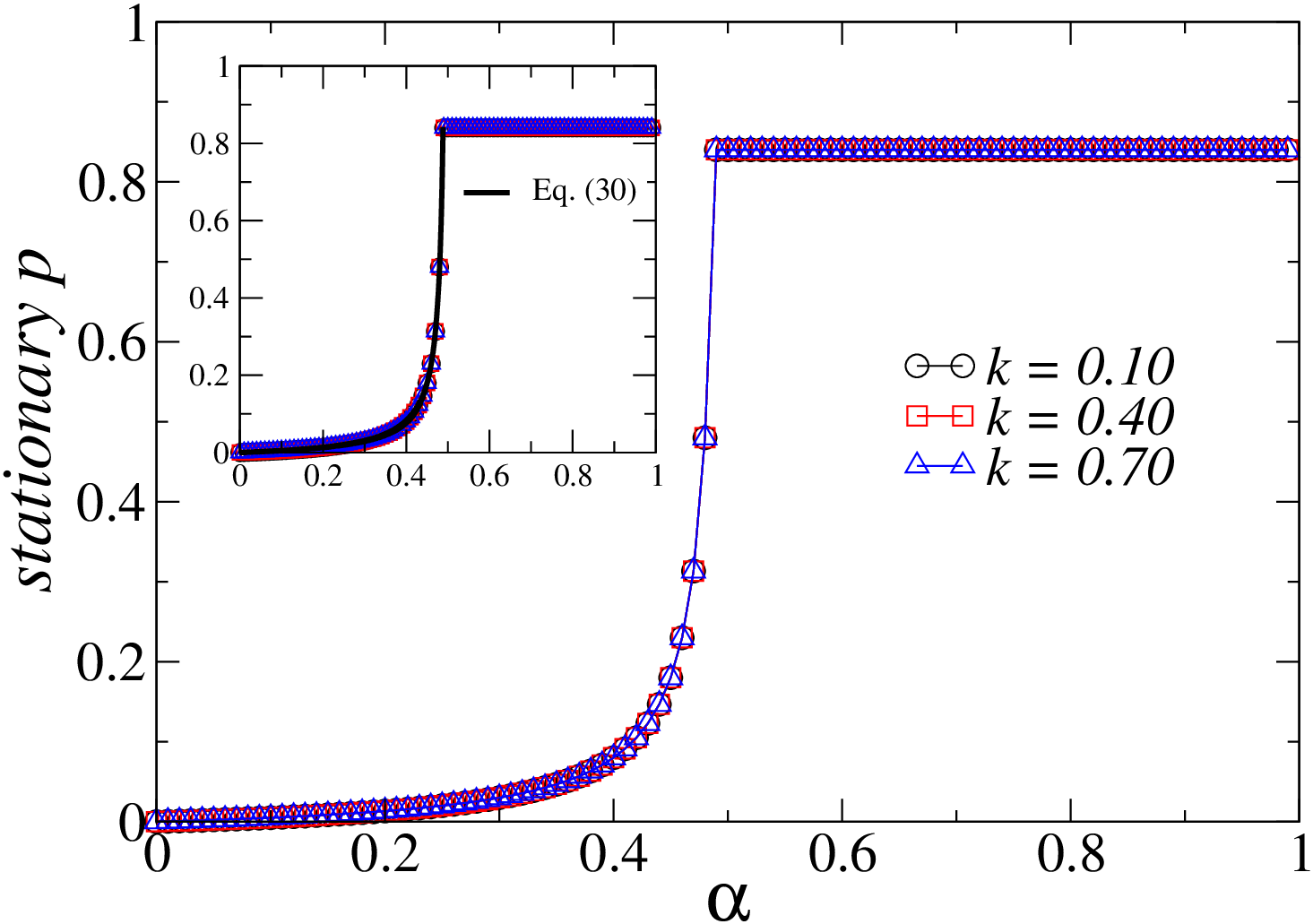}
\\
\vspace{1.0cm}
\includegraphics[width=0.48\textwidth,angle=0]{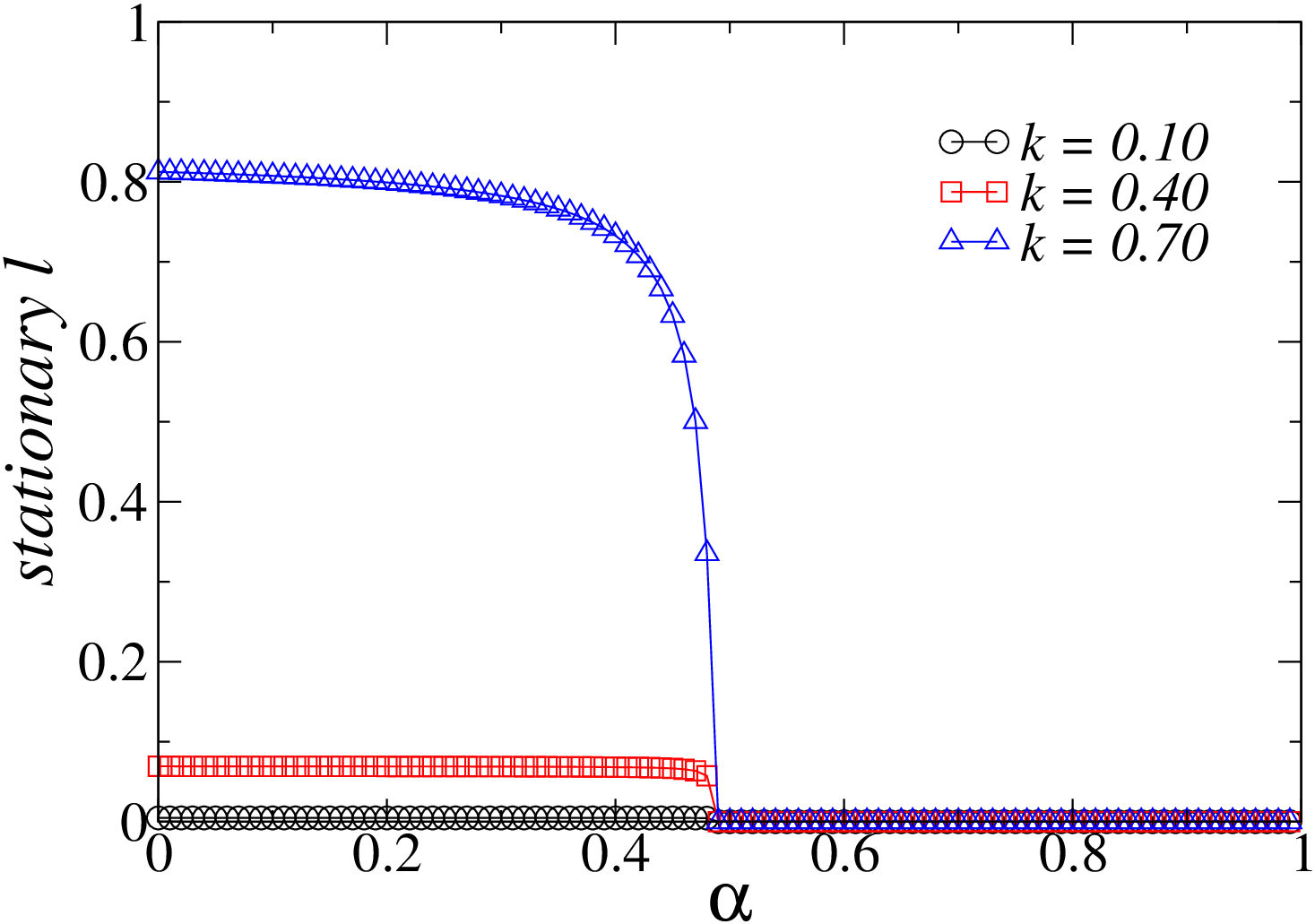}
\includegraphics[width=0.48\textwidth,angle=0]{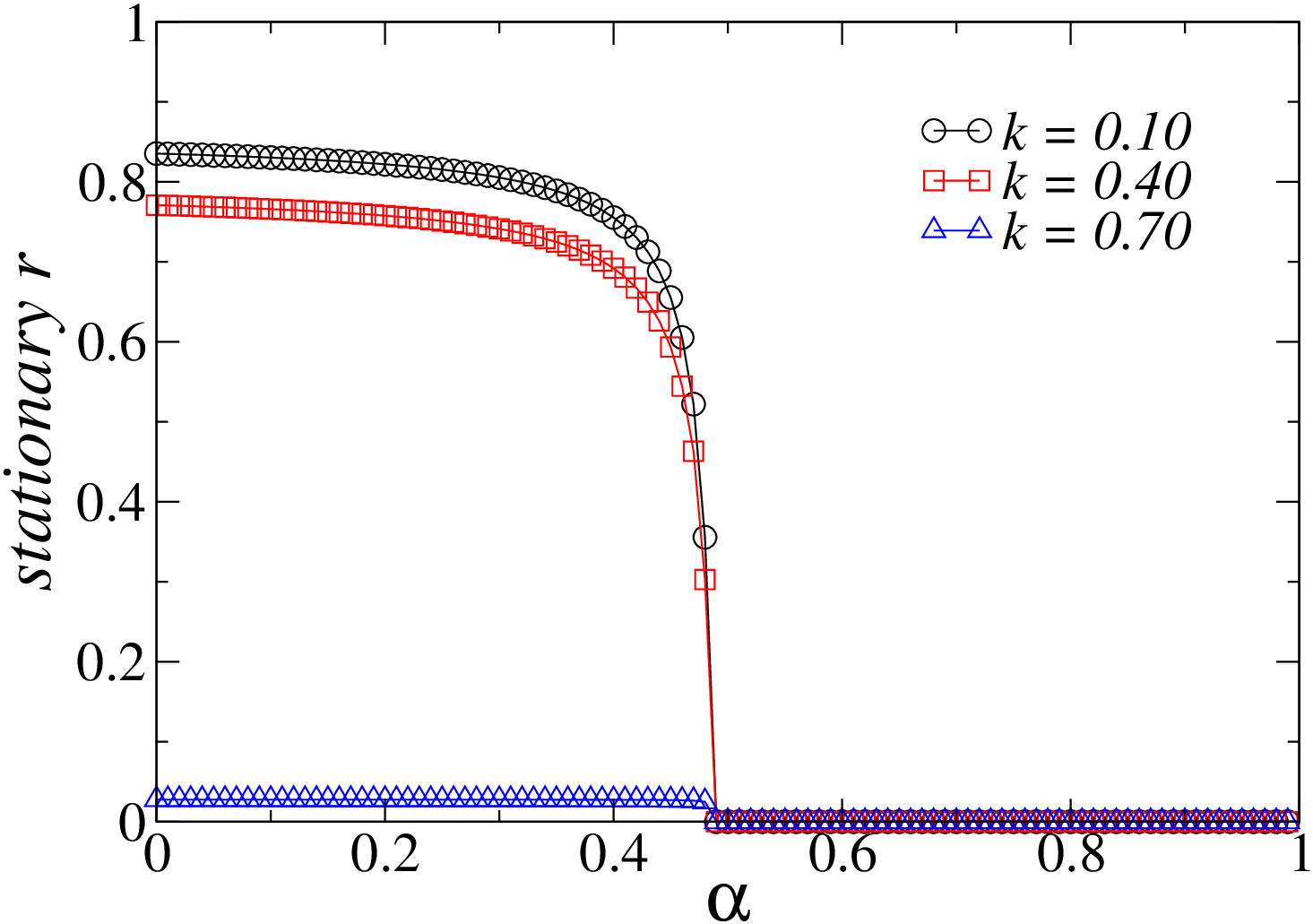}
\end{center}
\caption{(Color online) (a: upper left)  Values of $k$ matching the elections results with no populism. The values with populism are added for 2018 and 2021. The x-axis is the time (in years) since the first election: $0$ represents the 1989 election, $5$ represents the 1994 election, and so on. (b: upper right, c: lower left, d: lower right) The plots show the stationary densities of populists $p$, left sensitives $l$ and right sensitives $r$ as a functions of $\alpha$ with $f=0.04$. The symbols are numerical results, and the lines are just guides to the eye. In panel (b), the inset exhibits the same data for $p$ only with symbols (circles, squares and triangles), together with the analytical result of Eq. \eqref{eq_p} (full black line).}
\label{fig3}
\end{figure}

In Table \ref{Tab2} we exhibit the values of $k$ considered to match the 2018 election as well as the 2021 poll. For comparison, we included in Tab. \ref{Tab2} the values of $k$ obtained for the model without populists. For $\alpha=0.30$ we get respectively $k=0.496$ and $k=0.520$ (see Table \ref{Tab2}) and the same equilibrium value of populists $p=0.03$. When $\alpha=0.40$ we found $k=0.499$ and $k=0.528$ and still a unique $p=0.08$ (see Table \ref{Tab2}). All the results are exhibited Figure \ref{fig3} (a, b).

These small values of  the proportions of populists are unexpected and puzzling when dealing with a candidate pictured as populist. To check this ``anomaly" we look at what happens increasing the populist coupling to $\alpha=0.50$. The increase of only $0.10$ reveals a sudden sharp jump in the equilibrium value of populists at $p=0.84$ against $p=0.08$ for $\alpha=0.40$.

Above jump in the value of populists hints at a singularity at some value of the populist coupling. To identify this singularity we plot in Figure \ref{fig3} (b, c, d) the variations of stationary fractions of populists $p$, left sensitives $l$ and right sensitives $r$ as a function of $\alpha$ for three different values $k=0.10, 0.40, 0.70$. Indeed, abrupt quasi-vertical jump and falls are seen to occur at $\alpha=0.50$.

In particular, Figure \ref{fig3} (b) shows that the increase of $p$ is very slow and stays at low values till the vicinity of $\alpha=0.45$, before a sudden jump to each high values. This behavior sheds a new light on the phenomenon of populism, which could explain why it is almost impossible to oppose it when it bursts.  Indeed, since the impact of populism is minor on the opinion dynamics during about half the range of the associated parameters, it generates a low proportional reaction, which in turn is overwhelmed by  the tidal wave populism when it reaches its breaking point.

These numerical results regarding the stationary fraction of populists only as a function of $\alpha$ can be understood if we inspect Eq. \eqref{eq31}. In the stationary state we have $dp/dt=0$, and Eq. \eqref{eq31} leads to the expression $[(1-2\alpha)\,(l+r)]p = \alpha\,f\,(l+r)$. Thus, we obtain
\begin{equation} \label{eq_p}
p = p(\alpha) = f\,\frac{\alpha}{1-2\alpha}  ~.
\end{equation}
This result is valid for $\alpha<0.50$, $l\neq 0$ and $r\neq 0$. Eq. \eqref{eq_p} shows that the stationary values of $p$ does not depend on $k$, as we discussed before. Indeed, $p$ is solely a function of $\alpha$, and the function $p(\alpha)$ has a jump at $\alpha=0.50$. Eq. \eqref{eq_p} explains our previous observations, and it is exhibited in the inset of Figure \ref{fig3} (b) together with the numerical results for the stationary $p$ obtained from the numerical integration of Eqs. \eqref{eq29} - \eqref{eq31}. In addition, as we observed in Figure \ref{fig3} (c,d) that $l=r=0$ for $\alpha>0.50$, from the normalization condition \eqref{eq20} we have $p=1-4f$. For the case of Figure \ref{fig3} (b), we have $f=0.04$ and thus $p=0.84$ for $\alpha>0.50$.

Above results lead to question the actual narrative which explains Bolsonaro 2018 victory as a populist event. At this stage, our analysis yields two possible scenarios with very different underlying political mechanisms:

\begin{enumerate}
\item Bolsonaro 2018 victory was not the result of populism but the outcome of the regular competition between sensitives from left and right.

\item Bolsonaro did benefit from populism but at a very low level with only a few percent.
\end{enumerate}

Such discussion can be connected with a recent work discussing whether Bolsonaro's victory is a remarkable case or a new effective model of public policy and political leadership for Brazil and other countries in the region \cite{populism_new_paper}. The author concluded that it will depend largely on Bolsonaro's success as the President of Brazil: at the moment, the interim results of his presidency are highly controversial and rather continue to divide and polarize Brazilian society.

The results of our mathematical model agree with a recent analysis performed by social scientists. Based on results from surveys, the authors concluded that support for the far-right candidate Jair Bolsonaro is explained by right-wing ideology and illiberal attitudes, with populist attitudes playing a small role, if any \cite{castanho2022}.


\section{Anticipating the 2022 election}

In case above second scenario is valid, to anticipate the 2022 outcome we need to stress that as long as $\alpha<0.50$, the equilibrium value of $p$ stays low and what matters is the value of $k$. However, as soon as $\alpha\geq 0.50$ the value of $k$ becomes irrelevant, and in such a case the populism would contribute strongly to Bolsonaro making likely his victory in 2022.

That would make the 2021 poll misleading to forecast the 2022 outcome since it was implemented with low populism couplings putting Bolsonaro in a status of an hidden winner since during the 2022 campaign these couplings could be increased moderately to reach the jump of $p$ as seen from Figure \ref{fig3} (b) and then propels Bolsonaro to victory. That would result in a surprising reversing of the 2021 Lula tilde like poll outcome in 2022 at the benefit of Bolsonaro.


\section{Final Remarks}

In this work, using a contagious-like model we have studied the dynamics of opinion to describe the evolution of the political voting trends in Brazil after the dictatorship from 1985 to nowadays. The model articulates around a system of coupled ordinary differential equations to account for the interactions of left and right supporters. For each side we have considered three kinds of agents, sensitives, inflexibles and radicals where only sensitives can shift opinions.

The central question we addressed in our manuscript was wether it is necessary to introduce a new type of populist political orientation to yield 2018 Bolsonaro victory as claimed by many analyses after his election \cite{,castanho2022,prerna}. Our results have shown that this condition is not necessary to yield Bolsonaro victory. Moreover, we found that if despite this finding we do introduce a new populist component, its impact on the 2018 outcome was marginal. However in this case, this marginal component could render possible a second Bolsonaro victory in 2022.

Such conclusions were obtained using two distinct settings of the model. First, we have assumed that radicals do not interact. They are kept apart from sensitives and inflexibles. The evolution of vote intentions are then ruled by interactions among sensitives and inflexibles, as well as among sensitives themselves. The model was shown to reproduce all elections outcomes after Brazilian dictatorship (1989 - 2018) using one single parameter $k$ with some fixed densities of inflexibles and radicals, that we considered equal for simplicity.

In a second step we have activated interactions of radicals with sensitives to create populists from sensitives.  Once created, populists interact with sensitives to accelerate their spreading.

Regarding the 2018 election, along the first approach driven solely by $k$, we showed that we can recover the 2018 outcome as well as the 2021 poll. Our hypothesis would mean that despite Bolsonaro populist narrative \cite{bolsonaro_populist}, his victory was the result of the same mechanisms, which have been at play during all past presidential elections. 

In our study, we found two scenarios of very different political contents, which could have equally led to Bolsonaro victory. While this finding has no effect on the the past 2018 election, it could put at stake the current belief that Bolsonaro will lose the 2022 election in case the second scenario is valid. Moreover, if it is the first scenario, according to our findings, the 2021 poll has overestimated Lula outcome.

Indeed, the two possible scenarios for 2018, with and without populists, were discussed in a recent paper by social scientists \cite{castanho2022}. They tested some hypothesis about the effects of populist attitudes and ideology on vote choice in 2018 election. The results suggest that the support for the far right populist candidate Bolsonaro is explained by right-wing ideology and illiberal attitudes, with populist attitudes playing a small role, if any. Their conclusions are in agreement with our findings that populism may not be the reason of past Bolsonaro victory and if it was, it contributed only at the margin. On this basis, we argue that a significant source of populism is potentially available for Bolsonaro at next election.


\section*{Acknowledgments}

NC acknowledges financial support from the Brazilian scientific funding agencies CNPq (Grant 310893/2020-8) and FAPERJ (Grant 203.217/2017). We also thank L. Sigaud for a critical reading of the manuscript.

\bibliographystyle{elsarticle-num-names}

\end{document}